%% file: main_EE.tex
\documentclass[journal,12pt,draftclsnofoot,onecolumn,twoside]{IEEEtran}

\usepackage{ifpdf}
 \ifpdf
     \usepackage[pdftex]{graphicx}
 \else
     \usepackage[dvips]{graphicx}
 \fi

\graphicspath{{figures/}}

\usepackage{cite}

\usepackage[cmex10]{amsmath}
\usepackage{amssymb,amsfonts}

\usepackage{algorithm}
\usepackage{algpseudocode}

\usepackage{times}

\usepackage{flushend}

\begin{document}
\title{Energy-Efficient Power Loading for OFDM-based Cognitive Radio Systems with Channel Uncertainties}

\author{
{Ebrahim Bedeer,
Osama Amin,
Octavia A. Dobre,
Mohamed H. Ahmed, and
Kareem E. Baddour,
}\\
}

\maketitle

\input{abstract}



%

\input{intro}
\input{model}

\input{opt}

\input{proposed}
\input{sim}

\input{conc}




%

\section*{Acknowledgment}
The authors would like to thank the Editor and the anonymous reviewers for their feedback which helped us to considerably improve the quality of the manuscript.


\end{document}

%% file: abstract.tex
\begin{abstract}
In this paper, we propose a novel algorithm to optimize the energy-efficiency (EE) of orthogonal frequency division multiplexing-based cognitive radio systems under channel uncertainties.
We formulate an optimization problem that guarantees a minimum required rate and a specified power budget for the secondary user (SU), while restricting the interference to primary users (PUs) in a statistical manner.
The optimization problem is non-convex and it is transformed to an equivalent problem using the concept of fractional programming. Unlike all related works in the literature, we consider the effect of imperfect channel-state-information (CSI) on the links between the SU transmitter and receiver pairs and we additionally consider the effect of limited sensing capabilities of the SU.  Since the interference constraints are met statistically, the SU transmitter does not require perfect CSI feedback from the PUs receivers. Simulation results show that the EE deteriorates as the channel estimation error increases. Comparisons with relevant works from the literature show that the interference thresholds at the PUs receivers can be severely exceeded and the EE is slightly deteriorated if the SU does not account for spectrum sensing errors.
\end{abstract} 

%% file: intro.tex
\section{Introduction}
\IEEEPARstart{C}{ognitive} radio (CR) can considerably enhance the spectrum utilization efficiency by dynamically sharing the spectrum between licensed/primary users (PUs) and unlicensed/secondary users (SUs) \cite{cabric2008addressing}.
This is achieved by granting  SUs opportunistic access to the white spaces within  PUs spectrum, while controlling the interference to  PUs. Orthogonal frequency division multiplexing (OFDM) is recognized as an attractive modulation technique for CR due to its spectrum shaping flexibility, adaptivity in allocating vacant radio resources, and capability of analyzing the spectral activities of PUs \cite{wang2011new}. Generally speaking, the interference introduced to  PUs bands in OFDM-based CR networks can be classified as: 1) mutual interference (co-channel interference (CCI) and adjacent channel interference (ACI)) between the SU and PUs due to the non-orthogonality of their respective transmissions \cite{wang2011new} and 2) interference due to the SU's imperfect spectrum sensing capabilities \cite{cabric2008addressing}.

Most of the existing research has focused on optimizing the transmission rate of SUs while limiting the interference introduced to PUs to predefined thresholds (see, e.g., \cite{almalfouh2011interference, bansal2011adaptive} and references therein). Recently, optimizing the energy-efficiency (EE)---defined as the total energy consumed to deliver one bit, or its inverse\footnote{The EE can be defined as the number of bits per unit energy. However, it is common to define it as the total energy consumed to deliver one bit; please, see \cite{amin2012cooperative, wang2012optimal,  oto2012energy}.}
---has received increasing attention due to steadily rising energy costs and environmental concerns \cite{amin2012cooperative, wang2012optimal,  oto2012energy, xie2012energy, wangenergy, mao2013energy, mao2013energy2, amin2012opportunistic}.
Wang   \textit{et al.} in \cite{wang2012optimal} optimized the EE of an OFDM-based CR network subject to power budget and interference constraints; however, this comes at the expense of deteriorating the rate of the SU.
Oto \textit{et al.} in \cite{oto2012energy} found the optimal packet size that maximizes the EE of CR sensor networks while maintaining acceptable interference levels to the licensed PUs. In \cite{xie2012energy}, Xie \textit{et al.} investigated the problem of maximizing the EE of heterogeneous cognitive radio networks coexisting with femtocells.
Wang \textit{et al.} in \cite{wangenergy} optimized the EE of OFDM-based CR system subject to PUs interference constraints and different SUs rates.
In \cite{mao2013energy}, Mao \textit{et al.} optimized the EE of CR MIMO broadcast channels while guaranteeing certain interference threshold at the PUs receivers.
The same authors optimized the EE of OFDM-based CR systems subject to controlled interference leakage to PUs in \cite{mao2013energy2}.
To the authors' knowledge, all prior research on optimizing the EE has assumed that the SU has perfect spectrum sensing capabilities and perfect channel-state-information (CSI) for the links between the SU transmitter and receiver pairs \cite{mao2013energy, mao2013energy2, wang2012optimal, oto2012energy, xie2012energy, wangenergy}. However, in practice sensing is not fully reliable due to SU hardware limitations and variable channel conditions. Furthermore, it is also of practical importance to study the impact of channel estimation errors for the SU links on the EE optimization problem.


In this paper,
we formulate a novel EE optimization problem for the SU subject to its total transmit power budget and predefined quality-of-service (QoS) in terms of the minimum supported rate, as well as statistical constraints on the CCI and ACI to existing PUs. The optimization problem considers channel estimation errors for the links between the SU transmitter and receiver pairs, along with SU spectrum sensing errors. Furthermore, the SU does not rely on perfect CSI for the links between the SU transmitter and PUs receivers, since the interference constraints are met statistically.


The remainder of the paper is organized as follows. Section \ref{sec:model} introduces the system model. Section \ref{sec:opt} analyzes the optimization problem and outlines the proposed algorithm for its solution. Simulation results are presented in Section \ref{sec:sim}, while conclusions are drawn in Section \ref{sec:conc}.


%% file: model.tex
\section{System Model} \label{sec:model}
\subsection{System description}


The available spectrum is assumed to be divided into $L$ subchannels that are licensed to $L$ PUs. 
We assume that the SU periodically senses the PUs spectrum in order to identify vacant bands for its transmission. Without loss of generality, we consider that the SU senses that subchannel $m$, of bandwidth $B$, is vacant.
However, due to the varying channel conditions between the SU and PUs, the SU may not detect the presence of the $m$th PU.
This means that the SU identifies the $m$th PU band as vacant when it is truly occupied. This is referred to as a mis-detection error and it is assumed to occur with a probability $\rho^{(m)}_{\rm{md}}$. On the other hand, the SU may identify the $\ell$th PU band as occupied when it is truly vacant. This is referred to as a false-alarm error and it is assumed to occur with a probability $\rho^{(\ell)}_{\rm{fa}}$.  Mis-detection errors lead to severe co-channel interference to the $m$th PU, while false-alarm errors result in the SU wasting transmission opportunities.

\subsection{Modeling the statistical CCI and ACI constraints with imperfect SU sensing}
Using the Bayes' theorem and the law of total probability, the probability that subchannel $m$ is truly occupied under the condition that the SU identified it to be vacant can be expressed as \cite{almalfouh2011interference}
\begin{IEEEeqnarray}{c}
\beta_{\rm{ov}}^{(m)} = \frac{\rho^{(m)}_{\rm{md}}  \rho^{(m)}}{\rho^{(m)}_{\rm{md}}  \rho^{(m)} + (1 - \rho^{(m)}_{\rm{fa}}) (1 - \rho^{(m)})}, \label{eq:b_ov}
\end{IEEEeqnarray}
where $\rho^{(m)}$ is the probability that the PU transmits on subchannel $m$ and $\beta_{\rm{ov}}^{(m)}$ represents the probability that the interference due to mis-detection errors will be present in subchannel $m$, which is determined to be vacant by the SU.
Furthermore, the probability that subchannel $\ell$ is truly occupied by the PU under the condition that the SU identified it to be occupied can be written as
\begin{IEEEeqnarray}{c}
\beta_{\rm{oo}}^{(\ell)} = \frac{ (1 - \rho^{(\ell)}_{\rm{md}})  \rho^{(\ell)}}{ (1 - \rho^{(\ell)}_{\rm{md}})  \rho^{(\ell)} + \rho^{(\ell)}_{\rm{fa}} (1 - \rho^{(\ell)})}. \label{eq:b_oo}
\end{IEEEeqnarray}
Note that for perfect sensing
$\beta_{\rm{ov}}^{(m)} = 0$ and $\beta_{\rm{oo}}^{(\ell)} = 1$.

Estimating the channel gains between the SU transmitter and the PUs receivers is challenging without the PUs cooperation. Hence, we assume that the SU transmitter has only knowledge of the fading distribution type and its corresponding parameters of the channels on these links. This is a reasonable assumption for certain wireless environments. For example, a Rayleigh distribution is usually assumed for the magnitude of the fading channel coefficients in non-line-of-sight urban environments.
The constraint on the CCI from the SU to the $m$th PU is formulated as $\beta_{\rm{ov}}^{(m)} |\mathcal{H}_{\rm{sp}}^{(m)} |^2 G^{(m)}  \sum_{i = 1}^{N} p_i \leq P_{\rm{th}}^{(m)},$
where $\mathcal{H}_{\rm{sp}}^{(m)}$ and $G^{(m)}$ are the channel gain and the distance-based path loss\footnote{The SU is assumed to know the PUs location information by accessing a Radio Environment Map \cite{zhao2007applying}.} to the distant $m$th PU receiver, $p_i$ is the power allocated to subcarrier $i$, $i = 1, ..., N$, and $P_{\rm{th}}^{(m)}$ is the interference threshold at the $m$th PU receiver. Since $\mathcal{H}_{\rm{sp}}^{(m)}$ is not perfectly known at the SU transmitter, the CCI constraint is limited below the threshold $P_{\rm{th}}^{(m)}$ with at least a probability of $\Psi_{\rm{th}}^{(m)}$. This is formulated as $\textup{Pr}\left(\beta_{\rm{ov}}^{(m)} | \mathcal{H}_{\rm{sp}}^{(m)} |^2 G^{(m)}     \sum_{i = 1}^{N} p_i \leq P_{\rm{th}}^{(m)} \right) \geq \Psi_{\rm{th}}^{(m)}.$
A non-line-of-sight propagation environment is assumed; therefore, the channel gain $\mathcal{H}_{\rm{sp}}^{(m)}$ can be modeled as a zero-mean complex Gaussian random variable, and, hence, $| \mathcal{H}_{\rm{sp}}^{(m)} |^2$ follows the exponential distribution \cite{proakisdigital}. After some mathematical manipulations, the CCI statistical constraints  can be expressed as $\sum_{i = 1}^{N} p_i \leq \frac{1}{\beta_{\rm{ov}}^{(m)}} \frac{\nu^{(m)}}{G^{(m)} \left(-\ln(1 - \Psi_{\rm{th}}^{(m)})\right)} P_{\rm{th}}^{(m)}$,
where $\frac{1}{\nu^{(m)}}$ is the mean of the exponential distribution.
To further reflect the SU transmitter's power amplifier limitations and/or satisfy regulatory maximum power limits, the total SU transmit power is limited to a certain threshold $P_{\rm{th}}$ as $\sum_{i = 1}^{N} p_i \leq P_{\rm{th}}$. Therefore, the constraint on the SU total transmit power can be generalized as
\begin{IEEEeqnarray}{c} \label{eq:CCI_constraint}
\sum_{i = 1}^{N} p_i \leq \left[P_{\rm{th}},\frac{1}{\beta_{\rm{ov}}^{(m)}} \frac{\nu^{(m)}}{G^{(m)} \left(-\ln(1 - \Psi_{\rm{th}}^{(m)})\right)} P_{\rm{th}}^{(m)}\right]^-,
\end{IEEEeqnarray}
where $[x,y]^-$ represents $\min(x,y)$. The ACI is mainly due to the power spectral leakage of the SU subcarriers to the PUs receivers. This depends on the power allocated to each SU subcarrier and the spectral distance between the SU subcarriers and the PUs receivers. Similar to the CCI constraint, the statistical ACI constraint can be written as
\begin{IEEEeqnarray}{c} \label{eq:ACI_constraint}
\sum_{i = 1}^{N} p_i \: \varpi_i^{(\ell)} \leq \frac{1}{\beta_{\rm{oo}}^{(\ell)}} \frac{\nu^{(\ell)}}{G^{(\ell)} \left(-\ln(1 - \Psi_{\rm{th}}^{(\ell)})\right)} P_{\rm{th}}^{(\ell)}, \quad \ell = 1, ..., L, \nonumber\\
\end{IEEEeqnarray}
where $\frac{1}{\nu^{(\ell)}}$ and $G^{(\ell)}$ are the mean of the exponential distribution and the distance-based path loss to the $\ell$th PU and $\varpi_i^{(\ell)} = T_{\rm{s}} \: \int_{f_{i,\ell}-\frac{B_\ell}{2}}^{f_{i,\ell}+\frac{B_\ell}{2}} \textup{sinc}^2(T_{\rm{s}} f) \: df$, with $T_{\rm{s}}$ as the SU OFDM symbol duration, $f_{i,\ell}$ as the spectral distance between the SU subcarrier $i$ and the $\ell$th PU  frequency band, $B_{\ell}$ as the bandwidth of the $\ell$th PU, and $\textup{sinc}(x) = \frac{\sin(\pi x)}{\pi x}$.

\subsection{Modeling the imperfect CSI on the link between the SU transmitter and receiver}
Unlike all the previous works in the literature that assume perfect CSI for the links between the SU transmitter and receiver pairs \cite{mao2013energy, mao2013energy2, oto2012energy, xie2012energy, wangenergy, wang2012optimal}, we consider the effect of the channel estimation errors on these links.
The channel is assumed to change slowly and is modeled as a time-invariant finite impulse response system with order equal to $N_{\rm{ch}}$, ${\bf{h}} = \left[ {h(0),} \, h(1), \, \cdots , \, h(N_{\rm{ch}})  \right]^T$, where each channel tap is assumed to be complex Gaussian distributed with zero-mean and variance $\sigma _{h}^2$. To avoid the intersymbol interference, a cyclic prefix is added at the SU transmitter and removed at the receiver. The noise at the SU receiver is modeled as additive white Gaussian noise (AWGN) with zero mean and correlation matrix equal to $\sigma ^2 _n \bf{I}$, where $\bf{I}$ is the identity matrix. The training pilot symbols $\bf{b}$ are added to the precoded block, where the receiver knows the pilot pattern and estimates the channel using the linear minimum mean square estimator (LMMSE) as ${\bf{\hat h}} = \left( {\sigma ^2 _n {\bf{R}}_h^{ - 1}  + {\bf{B}}^H {\bf{B}}} \right)^{ - 1} {\bf{B}}^H \bf{x},$
where $\bf{x}$ is the received block and $\bf{B}$ is an $N \times (N_{\rm{ch}}+1)$ column wise circulant matrix with the first column equal to $\bf{x}$ \cite{ohno2004capacity}. The subchannel estimates are computed as \cite{ohno2004capacity} $[ \hat H(1), \, \hat H(W), \dots , \hat H(W^{N-1})]^T = \sqrt{N} {\bf{F}}_{N_{\rm{ch}}} \bf{ \hat h},$
where $W=e^{j2\pi/N}$, ${\bf{F}}_{N_{\rm{ch}}}$ is a submatrix of $\bf{F}$ corresponding to the first $N_{\rm{ch}}+1$ columns, and ${\bf{F}}$ is the $N \times N$ discrete Fourier transform matrix with the $(l,n)$ element defined as $\left[ \bf{F} \right]_{l,n} = W^{-ln}/ \sqrt{N}$. The channel capacity is expressed in terms of the channel estimate across subcarriers \cite{ohno2004capacity}, while taking the interference from the PUs into account, as
\begin{equation}  \label{eq3}
c(\mathbf{p}) = \Delta f \sum\limits_{i = 1}^N {\log _2 \left( {1 + \frac{{\left| {\hat H\left( {W^i } \right)} \right|^2 G \, p_i }}{{\sigma _{\Delta H}^2 G \, p_i  + \sigma ^2 _n + \mathcal{J}_i}}} \right)},
\end{equation}
where $\Delta f$ is the subcarrier bandwidth, $\mathbf{p} = [p_1, ..., p_N]^{\rm{T}}$ is the vector representing the power allocated to each subcarrier, $G$ is the distance-based path loss, $\mathcal{J}_i$ is the interference from the PUs to subcarrier $i$ of the SU (it depends on the SU receiver windowing function and power spectral density of the PUs \cite{weiss2004mutual}), and $\sigma _{\Delta H}^2$ is the minimum mean square error (MMSE) of the channel estimate. The latter can be expressed as $\sigma _{\Delta H}^2  = ({\left( {N_{\rm{ch}} + 1} \right)\sigma _h^2  \sigma ^2 _n })/({\sigma _n^2   +  \sigma _h^2 G P_{\rm{pilots}} })$,
where $P_{\rm{pilots}}$ is the pilots' transmitted power \cite{ohno2004capacity}.

%% file: opt.tex
\section{Optimization Problem and Proposed Algorithm} \label{sec:opt}

\subsection{Optimization problem formulation and analysis}
Our target is to optimize the SU EE, under channel uncertainties, while guaranteeing a total transmit power budget, limiting the CCI and ACI to the $m$th and $\ell$th PUs receivers below certain thresholds with a predefined probability, and ensuring the SU QoS in terms of a minimum supported rate. In this paper, we minimize the EE defined as the total energy consumed to deliver one bit.
Accordingly, the optimization problem is formulated as
\begin{IEEEeqnarray}{c} \label{eq:MOOP_1}
\mathcal{OP}1: \quad \underset{p_i}{\textup{min}} \quad \eta_{\rm{EE}} =  \frac{\kappa \sum_{i = 1}^{N} p_i + p_{\rm{c}}} {c(\mathbf{p})} \nonumber \\
\textup{subject to} \qquad \textup{C1}: (\ref{eq:CCI_constraint}), \quad \textup{C2}: (\ref{eq:ACI_constraint}), \quad \textup{C3}: c(\mathbf{p}) \geq R_{\rm{th}},
\end{IEEEeqnarray}
where $\kappa$ is a constant that depends on the power amplifier efficiency, $p_{\rm{c}}$ is the circuitry  power consumption, and $R_{\rm{th}}$ is the minimum required SU rate. The objective function in (\ref{eq:MOOP_1}) is non-convex; hence, $\mathcal{OP}1$ is non-convex and the global optimal solution is not guaranteed. The non-convex optimization problem in (\ref{eq:MOOP_1}) can be transformed to an equivalent optimization problem using the concept of fractional programming \cite{dinkelbach1967nonlinear}. Let us define a new objective function as
\begin{IEEEeqnarray}{c}
\Phi(\mathbf{p}, q) = \kappa \sum_{i = 1}^{N} p_i + p_{\rm{c}} - q \: c(\mathbf{p}),
\end{IEEEeqnarray}
where $q$ is a non-negative parameter/constant (and not a variable). We define a new optimization problem $\mathcal{OP}2$ as
\begin{IEEEeqnarray}{c}
\mathcal{OP}2: \quad \underset{p_i}{\textup{min}} \quad  \Phi(\mathbf{p}, q), \qquad
\textup{subject to} \hspace{20pt} \textup{C1---C3}. \label{eq:OP1}
\end{IEEEeqnarray}
It was shown in \cite{dinkelbach1967nonlinear} that at a certain value of the parameter $q$, denoted as $q^*$, the optimal solution of $\mathcal{OP}2$ is also the optimal solution to $\mathcal{OP}1$. Hence, finding the optimal power allocation $\mathbf{p}^*$ of $\mathcal{OP}1$ can be realized by finding the optimal power allocation $\mathbf{p}^*(q)$ of $\mathcal{OP}2$; then update the value of $q$ until it reaches $q^*$ \cite{dinkelbach1967nonlinear}.
Following \cite{dinkelbach1967nonlinear}, let us define $\Phi_{\rm{min}}(q) = \underset{p_i}{\min} \{ \Phi(\mathbf{p}, q) | \mathbf{p} \in \mathcal{S}\}$ to be the minimum of $ \Phi(\mathbf{p}, q)$, where $\mathcal{S}$ is the non-empty feasible region of $\mathcal{OP}1$ and $\mathcal{OP}2$ and $q^*$ is the minimum of $\eta_{\rm{EE}}(\mathbf{p})$, i.e., $q^* = \eta_{\rm{EE}}(\mathbf{p}^*) = (\kappa \sum_{i = 1}^{N} p_i^* + p_{\rm{c}})/(c(\mathbf{p}^*))$. If $\Phi_{\rm{min}}(q^*) = 0$, then the power that corresponds to $q^* = \eta_{\rm{EE}}(\mathbf{p}^*)$ is the optimal solution of $\mathcal{OP}1$ \cite{dinkelbach1967nonlinear}.
$\mathcal{OP}2$ can be solved by applying the Karush-Kuhn-Tucker (KKT) conditions \cite{Boyd2004convex},
where the Lagrangian function is expressed as
\begin{IEEEeqnarray}{RCL}
\mathcal{L}(\mathbf{p},\mathbf{y},\boldsymbol \lambda) & = & \kappa \sum_{i = 1}^{N} p_i + p_{\rm{c}} - q \: c(\mathbf{p}) \nonumber \\ & + &  \lambda_{1} \Big[\sum_{i = 1}^{N} p_i \nonumber \\ & &- \big[P_{\rm{th}},\frac{1}{\beta_{\rm{ov}}^{(m)}} \frac{\nu^{(m)}}{G^{(m)} \left(-\ln(1 - \Psi_{\rm{th}}^{(m)})\right)} P_{\rm{th}}^{(m)}\big]^- + y_{1}^2\Big] \nonumber \\
& + &  \sum_{\ell = 1}^{L} \lambda_{2}^{(\ell)} \Big[ \sum_{i = 1}^{N} p_i \: \varpi_i^{(\ell)} \nonumber \\ & & - \frac{1}{\beta_{\rm{oo}}^{(\ell)}} \frac{\nu^{(\ell)}}{G^{(\ell)} \left(-\ln(1 - \Psi_{\rm{th}}^{(\ell)})\right)} P_{\rm{th}}^{(\ell)} + y_{2}^{(\ell)^2} \Big] \nonumber \\ & + &  \lambda_{3} \left[R_{\rm{th}} -  c(\mathbf{p}) + y_{3}^2 \right],
\end{IEEEeqnarray}
where $\boldsymbol \lambda = [\lambda_{1}, \lambda_{2}^{(\ell)}, \lambda_{3}]^{\rm{T}}$ and $\mathbf{y} = [y_1^2, y_2^{{(\ell)}^2}, y_3^2]^{\rm{T}}$, $\ell = 1, ..., L$, are the vectors of the Lagrange multipliers and slack variables, respectively. A stationary point can be found when $\nabla \mathcal{L}(\mathbf{p},\mathbf{y},\boldsymbol \lambda) = 0$, which yields

\begin{subequations}
\begin{IEEEeqnarray}{l}
{\scriptstyle
\frac{\partial \mathcal{L}}{\partial p_i} =  \frac{- \frac{\Delta f}{\ln(2)} (q + \lambda_{3})  | {\hat H\left( {W^i } \right)} |^2 G\: ( \sigma_n^2 + \mathcal{J}_i)}{\sigma _{\Delta H}^2 G^2 (\sigma _{\Delta H}^2 + | {\hat H\left( {W^i } \right)}|^2) p_i^2 + G \: ( \sigma_n^2 + \mathcal{J}_i) (2 \sigma _{\Delta H}^2 + | {\hat H\left( {W^i } \right)} |^2) p_i + \: ( \sigma_n^2 + \mathcal{J}_i)^2}  }\nonumber \\
\qquad \qquad \qquad \qquad \qquad {+ \kappa  + \lambda_{1} + \sum_{\ell = 1}^{L} \lambda_{2}^{(\ell)} \varpi_i^{(\ell)} = 0,
} \label{eq:L_first}\\
\frac{\partial \mathcal{L}}{\partial \lambda_{1}} = \sum_{i = 1}^{N} p_i - \left[P_{\rm{th}},\frac{1}{\beta_{\rm{ov}}^{(m)}}
\frac{\nu^{(m)}}{G^{(m)} (-\ln(1 - \Psi_{\rm{th}}^{(m)}))} P_{\rm{th}}^{(m)}\right]^- \nonumber \\ \qquad \qquad \qquad \qquad \qquad + y_{1}^2 = 0, \\ \nonumber
\frac{\partial \mathcal{L}}{\partial \lambda_{2}^{(\ell)}} = \sum_{i = 1}^{N} p_i \: \varpi_i^{(\ell)} - \frac{1}{\beta_{\rm{oo}}^{(\ell)}} \frac{\nu^{(\ell)}}{G^{(\ell)} (-\ln(1 - \Psi_{\rm{th}}^{(\ell)}))} P_{\rm{th}}^{(\ell)} \nonumber \\ \qquad \qquad \qquad \qquad \qquad + y_{2}^{(\ell)^2} = 0, \\
\frac{\partial \mathcal{L}}{\partial \lambda_{3}} = R_{\rm{th}} -  c(\mathbf{p}) + y_{3}^2 = 0,\\
\frac{\partial \mathcal{L}}{\partial y_{1}} = 2 \lambda_{1} \, y_{1} = 0, \label{eq:OP_1_6}\\
\frac{\partial \mathcal{L}}{\partial y_{2}^{(\ell)}} = 2 \lambda_{2}^{(\ell)} \, y_{2}^{(\ell)} = 0, \label{eq:OP_1_7}\\
\frac{\partial \mathcal{L}}{\partial y_{3}} = 2 \lambda_{3} \, y_{3} = 0,
 \label{eq:L_end}
\end{IEEEeqnarray}
\end{subequations}
It can be seen that (\ref{eq:L_first})--(\ref{eq:L_end}) represent $N + 2 L + 4$ equations in the $N + 2 L + 4$ unknown components of the vectors $\mathbf{p}, \mathbf{y}$, and $\boldsymbol \lambda$. From (\ref{eq:L_first}), the optimal power allocation per subcarrier is given as
\begin{IEEEeqnarray}{l}
p_i^* = \Big[\chi_i \Big(-1 + \big( 1 - \bigg(\frac{( \sigma_n^2 + \mathcal{J}_i)}{G}  \nonumber \\ \qquad \qquad \qquad- \frac{\frac{\Delta f}{\ln(2)} (q + \lambda_{3}) | {\hat H\left( {W^i } \right)} |^2}{\kappa + \lambda_{1} + \sum_{\ell = 1}^{L} \lambda_{2}^{(\ell)} \varpi_i^{(\ell)}} \bigg) \nonumber \\ \qquad \qquad \qquad \frac{2}{\chi_i \big( 2\sigma _{\Delta H}^2  + | {\hat H\left( {W^i } \right)} |^2  \big)}\big)^{1/2} \Big)\Big]^+, \label{eq:optimal}
\end{IEEEeqnarray}
where $[x]^+$ represents $\max(0,x)$ and the value of $\chi_i$ is calculated as

$\chi_i = ({( \sigma_n^2 + \mathcal{J}_i) ( {2\sigma _{\Delta H}^2  + | {\hat H( {W^i } )} |^2 } )})/({2\sigma _{\Delta H}^2 ( {\sigma_{\Delta H}^2  + | {\hat H( {W^i } )} |^2 } )G}).$
 
In (\ref{eq:optimal}), the values of the Lagrangian multipliers $\lambda_{1}$, $\lambda_{2}^{(\ell)}$, and $\lambda_{3}$ are determined based on whether the constraints on the CCI/total transmit power, ACI, and rate are active or inactive, respectively (a constraint on the form $\Gamma(x) \leq \Gamma_{\rm{th}}$ is said to be inactive if $\Gamma(x) < \Gamma_{\rm{th}}$, while it is active if $\Gamma(x) = \Gamma_{\rm{th}}$).
Equation (\ref{eq:OP_1_6}) implies that either $\lambda_{1} = 0$ or $y_{1} = 0$, (\ref{eq:OP_1_7}) implies that either $\lambda_{2}^{(\ell)} = 0$ or $y_{2}^{(\ell)} = 0$, and (\ref{eq:L_end}) implies that either $\lambda_{3} = 0$ or $y_{3} = 0$. Hence, eight possible cases exist, as follows:

---\textit{Cases 1 \& 2}: setting $\lambda_{1} = 0$, $\lambda_{2}^{(\ell)} = 0$, and $\lambda_{3} = 0$ (case 1)/$y_{3} = 0$ (case 2) results in the optimal solution for inactive CCI/total transmit power constraint, inactive ACI constraints, and inactive/active rate constraint, respectively.


---\textit{Case 3 \& 4}: setting $y_{1} = 0$, $\lambda_{2}^{(\ell)} = 0$, and $\lambda_{3} = 0$ (case 3)/$y_{3} = 0$ (case 4) results in the optimal solution for active CCI/total transmit power constraint, inactive ACI constraint, and inactive/active rate constraint, respectively.


---\textit{Case 5 \& 6}: setting $\lambda_{1} = 0$, $y_{2}^{(\ell)} = 0$, and $\lambda_{3} = 0$ (case 5)/$y_{3} = 0$ (case 6) results in the optimal solution for inactive CCI/total transmit power constraint, active ACI constraint, and inactive/active rate constraint, respectively.


---\textit{Case 7 \& 8}: setting $y_{1} = 0$, $y_{2}^{(\ell)} = 0$, and $\lambda_{3} = 0$ (case 7)/$y_{3} = 0$ (case 8) results in the optimal solution for active CCI/total transmit power constraint, active ACI constraint, and inactive/active rate constraint, respectively.

%% file: proposed.tex
\subsection{Proposed algorithm and complexity analysis}

The proposed algorithm can be formally stated as follows:
\floatname{algorithm}{}
\begin{algorithm}
\renewcommand{\thealgorithm}{}
\caption{\textbf{Proposed Algorithm}}
\begin{algorithmic}[1]
\small
\State \textbf{INPUT} $P_{\rm{th}}$, $P_{\rm{th}}^{(m)}$, $P_{\rm{th}}^{(\ell)}$, $R_{\rm{th}}$, $\nu^{(m)}$, $\nu^{(\ell)}$, $G^{(m)}$, $G^{(\ell)}$, $\Psi_{\rm{th}}^{(m)}$, $\Psi_{\rm{th}}^{(\ell)}$, $\beta_{\rm{ov}}^{(m)}$, $\beta_{\rm{oo}}^{(\ell)}$, $G$, $\sigma _n^2$, $\hat H\left( {W^i }\right)$, $\sigma _{\Delta H}^2$, $\Delta f$, $N$, $\delta > 0$, $q = q_{\rm{initial}}$ and $\Phi_{\rm{min}} = -\infty$.
\While{$ \Phi_{\rm{min}}(q) < - \delta$}
\State - assume the optimal solution $p_i^*$ belongs to case 1, i.e., $\sum_{i = 1}^{N} p_i^* < \left[P_{\rm{th}},\frac{1}{\beta_{\rm{ov}}^{(m)}} \frac{\nu^{(m)}}{G^{(m)} \left(-\ln(1 - \Psi_{\rm{th}}^{(m)})\right)} P_{\rm{th}}^{(m)}\right]^-$, $\sum_{i = 1}^{N} p_i^* \: \varpi_i^{(\ell)} < \frac{1}{\beta_{\rm{oo}}^{(\ell)}} \frac{\nu^{(\ell)}}{G^{(\ell)} \left(-\ln(1 - \Psi_{\rm{th}}^{(\ell)})\right)} P_{\rm{th}}^{(\ell)}$, and $c(\mathbf{p}) > R_{\rm{th}}$. \label{step:1}
\State - find $p_i^*$ from (\ref{eq:optimal}) when $\lambda_{1} = \lambda_{2}^{(\ell)} = \lambda_{3} = 0$.
\If{in Step \ref{step:1}, the assumption on the CCI/total transmit power constraint is true, the assumption on the ACI constraint is true, and the assumption on the rate constraint is not true}.
\State - the optimal solution belongs to case 2, i.e., find non-negative $\lambda_{3}$ from (\ref{eq:optimal}) such that $c(\mathbf{p}) = R_{\rm{th}}$.
\State - if the assumption on the CCI/total transmit power and ACI constraints are violated, then $p_i^* = 0$.
\ElsIf{in Step \ref{step:1}, the assumption on the CCI/total transmit power constraint is not true, the assumption on the ACI constraint is true, and the assumption on the rate constraint is true}
\State - the optimal solution belongs to case 3, i.e., find non-negative $\lambda_{1}$ from (\ref{eq:optimal}) such that $\sum_{i = 1}^{N} p_i^* = \left[P_{\rm{th}},\frac{1}{\beta_{\rm{ov}}^{(m)}} \frac{\nu^{(m)}}{G^{(m)} \left(-\ln(1 - \Psi_{\rm{th}}^{(m)})\right)} P_{\rm{th}}^{(m)}\right]^-$.
\State - if the assumption on the rate constraint is violated, then $p_i^* = 0$.
\ElsIf{in Step \ref{step:1}, the assumption on the CCI/total transmit power constraint is not true, the assumption on the ACI constraint is true, and the assumption on the rate constraint is not true}
\State - the optimal solution belongs to case 4, i.e., find non-negative $\lambda_{1}$ and $\lambda_{3}$ from (\ref{eq:optimal}) such that $\sum_{i = 1}^{N} p_i^* = \left[P_{\rm{th}},\frac{1}{\beta_{\rm{ov}}^{(m)}} \frac{\nu^{(m)}}{G^{(m)} \left(-\ln(1 - \Psi_{\rm{th}}^{(m)})\right)} P_{\rm{th}}^{(m)}\right]^-$ and $c(\mathbf{p}) = R_{\rm{th}}$.
\State - if the assumption on the ACI constraint is violated, then $p_i^* = 0$.
\ElsIf{in Step \ref{step:1}, the assumption on the CCI/total transmit power constraint is true, the assumption on the ACI constraint is not true, and the assumption on the rate constraint is true}
\State - the optimal solution belongs to case 5, i.e., find non-negative $\lambda_{2}^{(\ell)}$ from (\ref{eq:optimal}) such that $\sum_{i = 1}^{N} p_i^* \: \varpi_i^{(\ell)} = \frac{1}{\beta_{\rm{oo}}^{(\ell)}} \frac{\nu^{(\ell)}}{G^{(\ell)} \left(-\ln(1 - \Psi_{\rm{th}}^{(\ell)})\right)} P_{\rm{th}}^{(\ell)}$.
\State - if the assumption on the rate constraint is violated, then $p_i^* = 0$.
\ElsIf{in Step \ref{step:1}, the assumption on the CCI/total transmit power constraint is true, the assumption on the ACI constraint is not true, and the assumption on the rate constraint is not true}
\State - the optimal solution belongs to case 6, i.e., find non-negative $\lambda_{2}^{(\ell)}$ and $\lambda_{3}$ from (\ref{eq:optimal}) such that $\sum_{i = 1}^{N} p_i^* \: \varpi_i^{(\ell)} = \frac{1}{\beta_{\rm{oo}}^{(\ell)}} \frac{\nu^{(\ell)}}{G^{(\ell)} \left(-\ln(1 - \Psi_{\rm{th}}^{(\ell)})\right)} P_{\rm{th}}^{(\ell)}$ and $c(\mathbf{p}) = R_{\rm{th}}$.
        \algstore{myalg}
  \end{algorithmic}
\end{algorithm}

\floatname{algorithm}{}
\begin{algorithm}
 \renewcommand{\thealgorithm}{}
  \caption{\textbf{Proposed Algorithm} (continued)}
  \begin{algorithmic}
      \algrestore{myalg}
      \small
\State - if the assumption on the CCI/total transmit power constraint is violated, then $p_i^* = 0$.
\ElsIf{in Step \ref{step:1}, the assumption on the CCI/total transmit power constraint is not true, the assumption on the ACI constraint is not true, and the assumption on the rate constraint is true}
\State - the optimal solution belongs to case 7, i.e., find non-negative $\lambda_{1}$ and $\lambda_{2}^{(\ell)}$  from (\ref{eq:optimal}) such that  $\sum_{i = 1}^{N} p_i^* = \left[P_{\rm{th}},\frac{1}{\beta_{\rm{ov}}^{(m)}} \frac{\nu^{(m)}}{G^{(m)} \left(-\ln(1 - \Psi_{\rm{th}}^{(m)})\right)} P_{\rm{th}}^{(m)}\right]^-$ and $\sum_{i = 1}^{N} p_i^* \: \varpi_i^{(\ell)} = \frac{1}{\beta_{\rm{oo}}^{(\ell)}} \frac{\nu^{(\ell)}}{G^{(\ell)} \left(-\ln(1 - \Psi_{\rm{th}}^{(\ell)})\right)} P_{\rm{th}}^{(\ell)}$.
\State - if the assumption on the rate constraint is violated, then $p_i^* = 0$.
\ElsIf{in Step \ref{step:1}, the assumption on the CCI/total transmit power constraint is not true, the assumption on the ACI constraint is not true, and the assumption on the rate constraint is not true}
\State - the optimal solution belongs to case 8, i.e., find non-negative $\lambda_{1}$, $\lambda_{2}^{(\ell)}$, and $\lambda_{3}$  from (\ref{eq:optimal}) such that  $\sum_{i = 1}^{N} p_i^* = \left[P_{\rm{th}},\frac{1}{\beta_{\rm{ov}}^{(m)}} \frac{\nu^{(m)}}{G^{(m)} \left(-\ln(1 - \Psi_{\rm{th}}^{(m)})\right)} P_{\rm{th}}^{(m)}\right]^-$, $\sum_{i = 1}^{N} p_i^* \: \varpi_i^{(\ell)} = \frac{1}{\beta_{\rm{oo}}^{(\ell)}} \frac{\nu^{(\ell)}}{G^{(\ell)} \left(-\ln(1 - \Psi_{\rm{th}}^{(\ell)})\right)} P_{\rm{th}}^{(\ell)}$, and $c(\mathbf{p}) = R_{\rm{th}}$.
\Else
\State - $p_i^* = 0$.
\EndIf
\State - update $\Phi_{\rm{min}}(q) = \underset{p_i}{\min} \{ \Phi(\mathbf{p}, q) \} | \mathbf{p} \in \mathcal{S}\}$
\State - Calculate $q = \frac{\kappa \sum_{i = 1}^{N} p_i^* + p_{\rm{c}}} {c(\mathbf{p})}$.
\EndWhile
\State \textbf{OUTPUT} $q^* = q$ and $p_i^*$, $i$ = 1, ..., $N$.
\end{algorithmic}
\end{algorithm}


Efficient algorithms are presented in \cite{palomar2005practical} to find the Lagrange multipliers $\lambda_{1}$ and $\lambda_{2}^{(\ell)}$, and $\lambda_{3}$ that satisfy the CCI/total transmit power, ACI, and rate constraints, respectively, with complexity order of $\mathcal{O}(N)$. Accordingly, the complexity order of the proposed algorithm can be $\mathcal{O}(N_qN^2)$, where $N_q$ is the number of executions of the while loop. The average (over the number of channel realizations) value for $N_q$ is 4 for $\delta = 10^{-8}$ and 4.46 for $\delta = 10^{-14}$; both values are significantly lower than the number of subcarriers $N$. Hence, the complexity of the proposed algorithm is of the order  $\mathcal{O}(N^2)$.


%% file: sim.tex
\section{Numerical Results} \label{sec:sim}

Without loss of generality, we assume that the OFDM SU coexists with one frequency-adjacent PU and one co-channel PU. The SU parameters are chosen as follows: number of subcarriers $N = 128$ and subcarrier spacing $\Delta f = \frac{1.25 \: \rm{MHz}}{N} = 9.7656$ kHz. The propagation path loss parameters are as follows: distance between SU transmitter and receiver pair $= 1$ km, distance to the $\ell$th PU $d_{\ell} = 1.2$ km, distance to the $m$th PU $d_m = 1.5$ km, reference distance $= 100$ m, exponent $= 4$, and wavelength $= \frac{3 \times 10^8}{900 \times 10^6} = 0.33$ meters. A Rayleigh fading environment is considered with $N_{\rm{ch}} = 5$, where the average channel power gains between the SU transmitter and the receiver of the $\ell$th PU $\mathbb{E}\{|\mathcal{H}_{sp}^{(\ell)}|^2\}$ and between the SU transmitter and the receiver of the $m$th PU $\mathbb{E}\{|\mathcal{H}_{sp}^{(m)}|^2\}$ are set to 0 dB. $\sigma_n^2$ is assumed to be $4 \times 10^{-16}$ W, the PUs signal is assumed to be an elliptically filtered white noise process \cite{weiss2004mutual} of variance $4 \times 10^{-16}$ W, $p_c = P_{\rm{th}} = 2$ W, $\kappa = 7.8$, $\delta = 10^{-8}$, $\Psi_{\rm{th}}^{(m)} = \Psi_{\rm{th}}^{(\ell)} = 0.9$, and $P_{\rm{th}}^{(m)} = P_{\rm{th}}^{(\ell)} = 10^{-13}$ W. Representative results are presented in this section, which were obtained through Monte Carlo trials for $10^{4}$ channel realizations. Unless otherwise mentioned, imperfect spectrum sensing is assumed. Following \cite{almalfouh2011interference} and in order to favor the PUs protection,  $\rho_{\rm{md}}^{(m)}$ is uniformly distributed over the interval [0.01,  0.05], and it is lower than $\rho_{\rm{fa}}^{(m)}$, which is uniformly distributed over the interval [0.01,  0.1].
$\rho^{(m)}$ and $\rho^{(\ell)}$ are uniformly distributed between [0, 1] and the EE, measured in J/bits, is the total energy consumption to deliver one bit.
\begin{figure}[!t]
\centering
\includegraphics[width=0.5\textwidth]{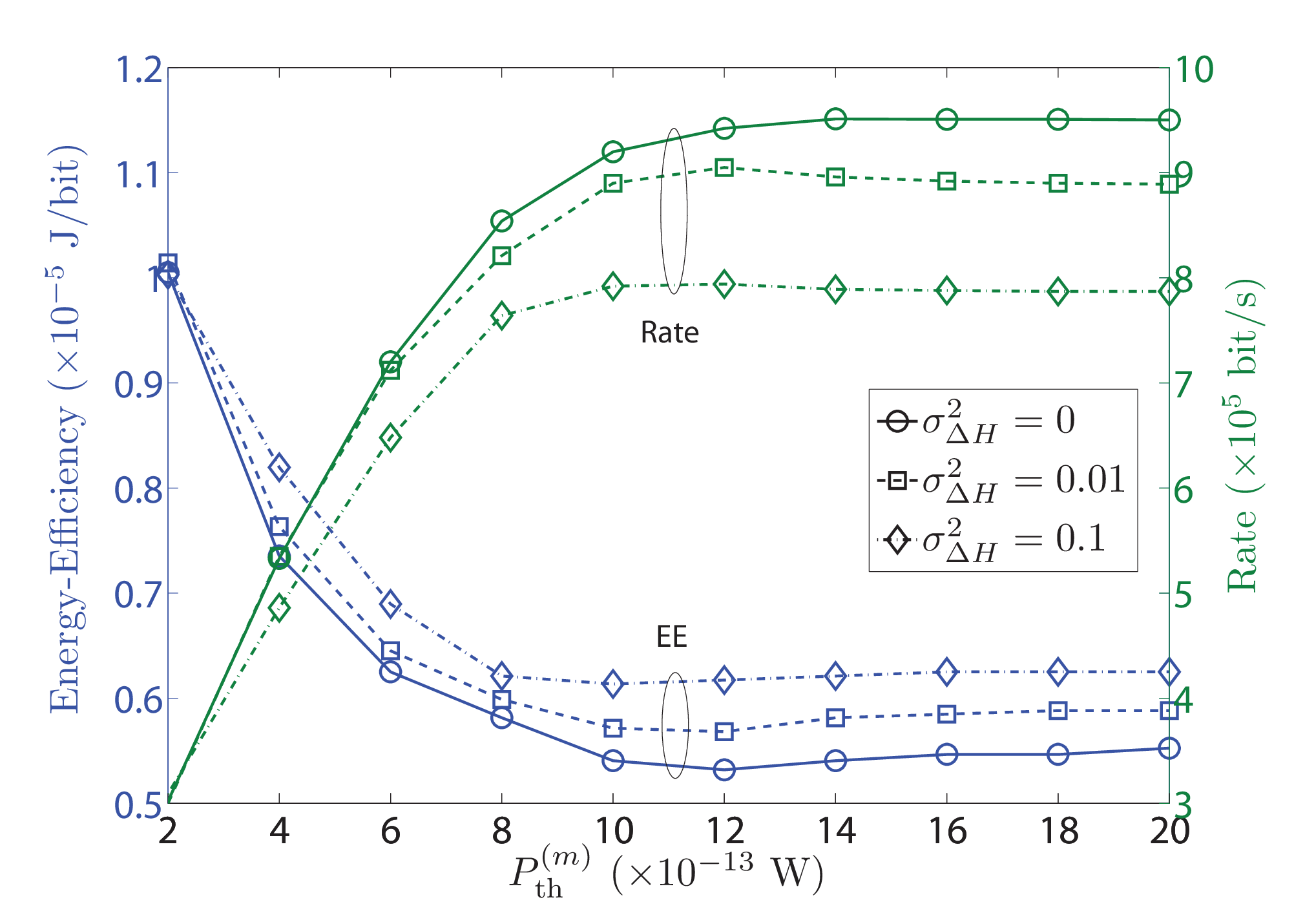}
\caption{Effect of $\sigma_{\Delta H}^2$ on the SU performance.}
\label{fig:MMSE}
\end{figure}

\begin{figure}[!t]
\centering
\includegraphics[width=0.5\textwidth]{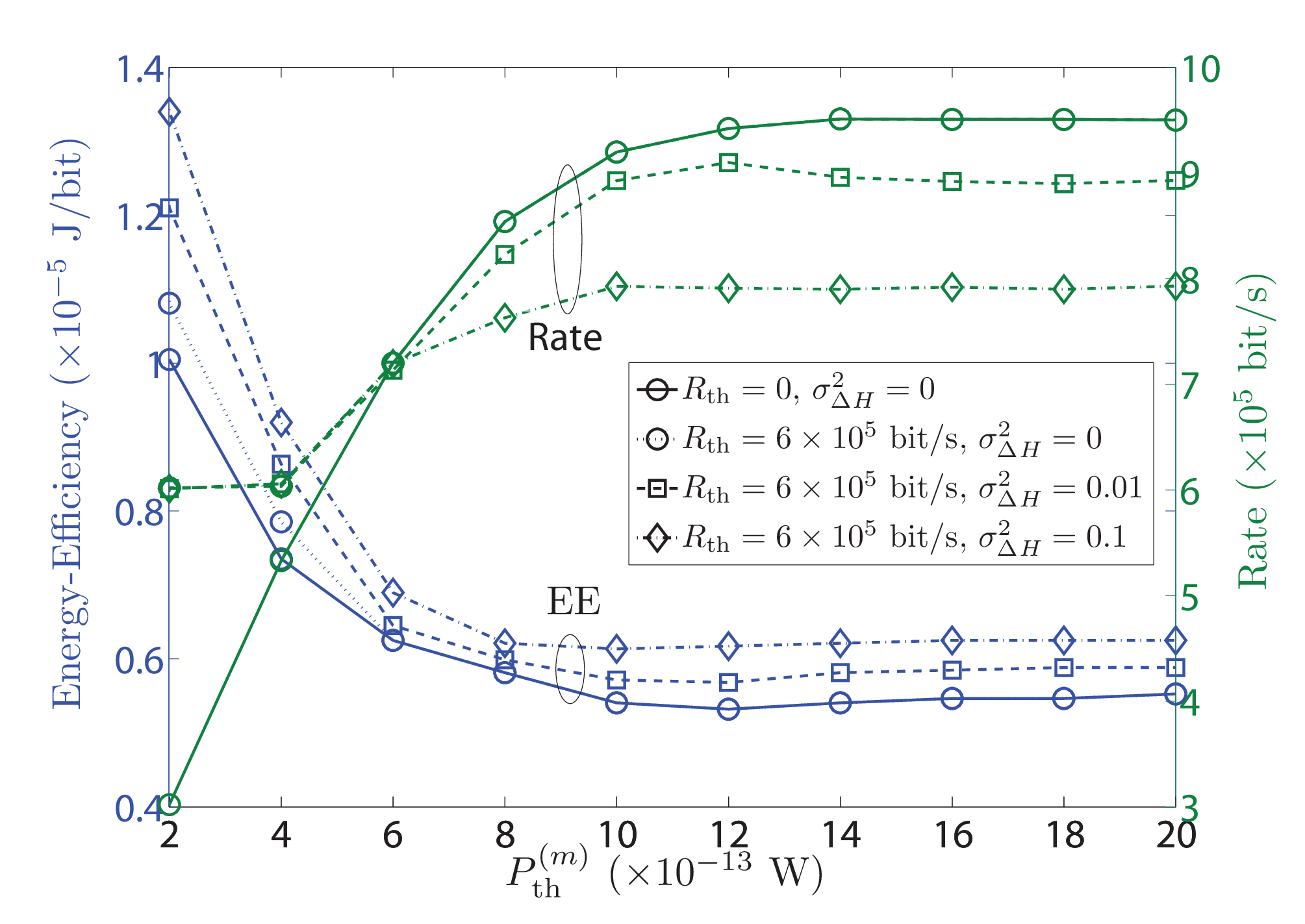}
\caption{Effect of $R_{\rm{th}}$ and $\sigma_{\Delta H}^2$ on the SU performance.}
\label{fig:PCCI}
\end{figure}


In Fig. \ref{fig:MMSE}, the EE (in J/bits) and the transmission rate (in bits/sec) of the SU  are depicted as a function of $P_{\rm{th}}^{(m)}$, for $R_{\rm{th}} = 0$ and different values of $\sigma_{\Delta H}^2$. As can be seen, the EE decreases and the rate increases as $P_{\rm{th}}^{(m)}$ increases, and both saturate for higher values of $P_{\rm{th}}^{(m)}$. This can be explained, as for lower values of $P_{\rm{th}}^{(m)}$ the total transmit power is limited, and increasing $P_{\rm{th}}^{(m)}$ increases the transmit power, and, hence, enables the proposed algorithm to improve both the EE and rate of the SU. The EE keeps improving until the optimal power budget is reached, after which a further increase in $P_{\rm{th}}^{(m)}$  does not improve the EE, and, hence, the rate is kept constant. As the value of $\sigma_{\Delta H}^2$ increases, i.e., the estimation error increases, both the EE and the rate  deteriorate accordingly.

Fig. \ref{fig:PCCI} depicts the SU EE and rate as a function of $P_{\rm{th}}^{(m)}$, for different values for $R_{\rm{th}}$ and $\sigma_{\Delta H}^2$. As expected, for $\sigma_{\Delta H}^2 = 0$, increasing $R_{\rm{th}}$ from 0 to $6 \times 10^5$ bits/sec guarantees the SU rate at low values of $P_{\rm{th}}^{(m)}$ (i.e., when the rate drops below $6 \times 10^5$ bits/sec); however, this comes at the expense of increasing the EE. On the other hand, for $R_{\rm{th}} = 6 \times 10^5$ bits/sec, increasing the estimation error deteriorates both the rate and the EE of the SU at high values of $P_{\rm{th}}^{(m)}$; for low values of the $P_{\rm{th}}^{(m)}$, the SU maintains its required rate but this at the expense of increasing the EE.

\begin{figure}[!t]
\centering
\includegraphics[width=0.50\textwidth]{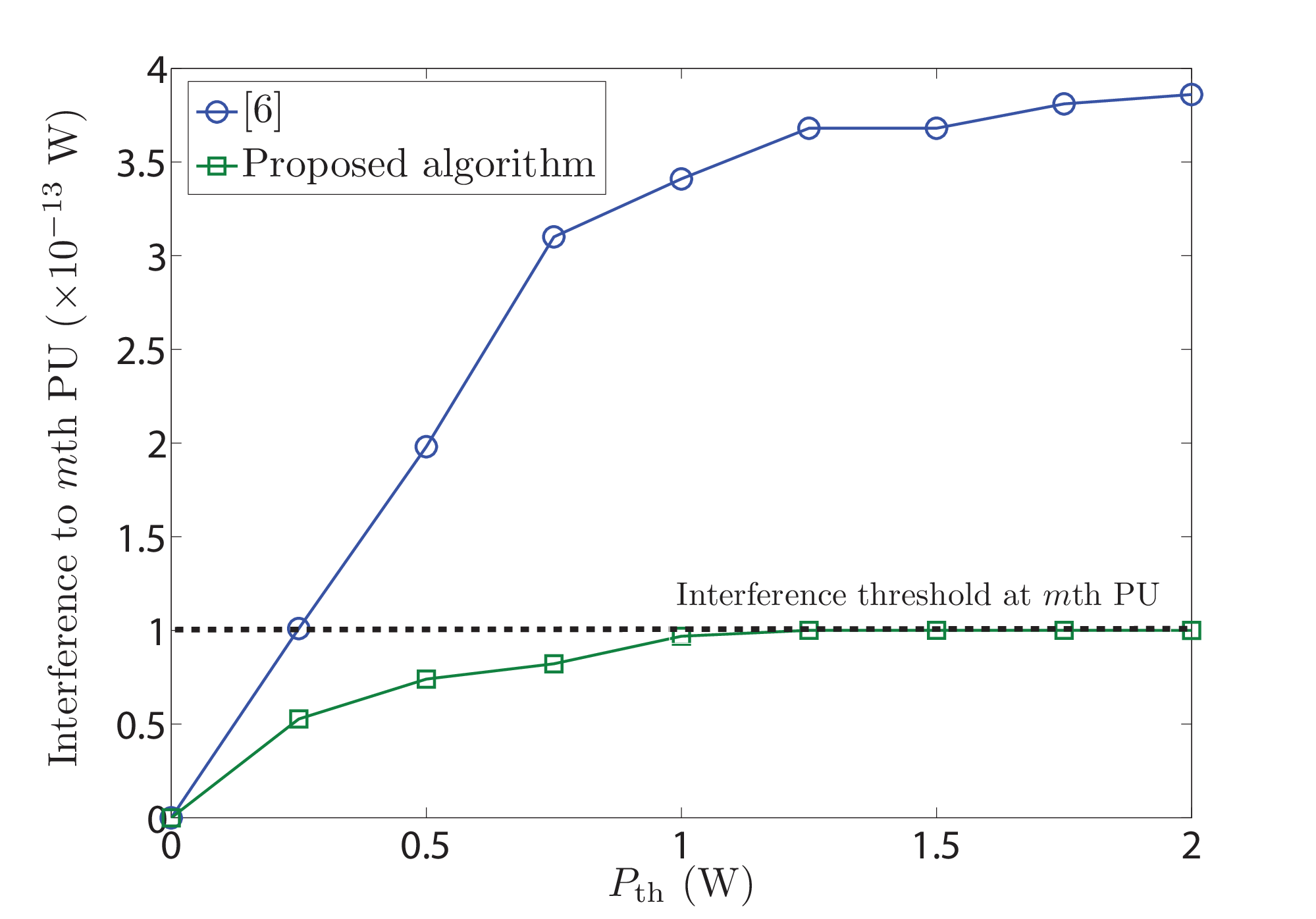}
\caption{Comparison with the work in \cite{wang2012optimal} to show the effect of perfect and imperfect sensing assumptions on the interference leaked to the $m$th PU.}
\label{fig:comp_interference}
\end{figure}
\begin{figure}[!t]
\centering
\includegraphics[width=0.50\textwidth]{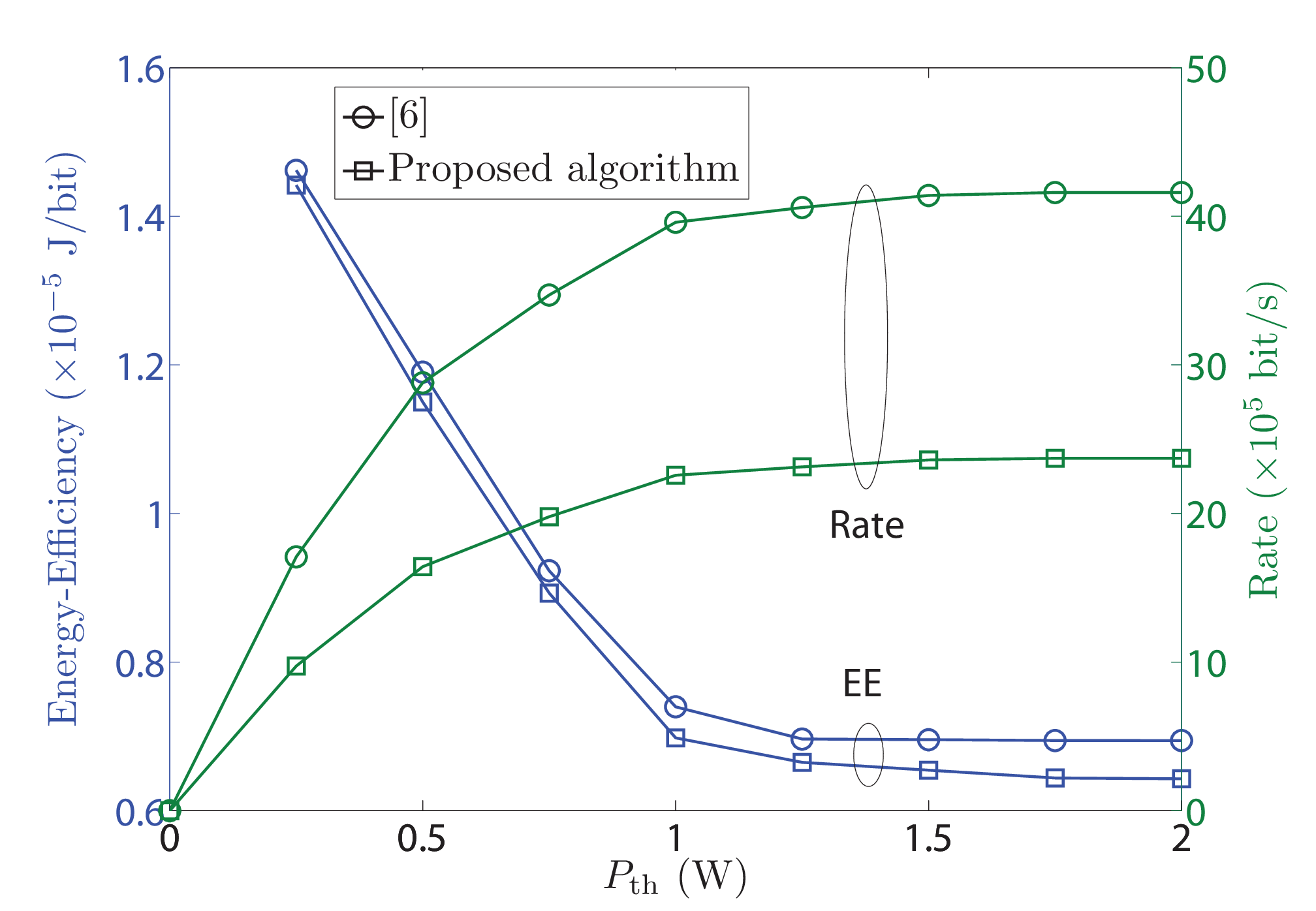}
\caption{Comparison with the work in \cite{wang2012optimal} to show the effect of perfect and imperfect sensing assumptions on the EE and the rate of SU.}
\label{fig:comp_EE_SE}
\end{figure}

In order to show the effect of assuming perfect spectrum sensing, Figs. \ref{fig:comp_interference} and \ref{fig:comp_EE_SE} compare the interference introduced into the $m$th PU band, and the EE and rate, respectively, for the proposed algorithm and the work in \cite{wang2012optimal} that assumes perfect sensing capabilities for the SU. We set $\sigma_{\Delta H}^2 = 0$ and $R_{\rm{th}} = 0$ in the proposed algorithm, in order to match the conditions in \cite{wang2012optimal}.
As can be seen in Fig. \ref{fig:comp_interference}, if the sensing errors are not taken into consideration when optimizing the EE as in \cite{wang2012optimal} (i.e., the SU is assumed to sense the PUs bands perfectly, which is not true in practice),
then the interference leaked in the $m$th PU band exceeds the threshold (note that this is due to the increase of the transmit power for the case of perfect spectrum sensing assumption). On the other hand, if the sensing errors are considered in the optimization problem (i.e., the SU is assumed to sense the PUs bands with a certain probability of error), then the interference to the $m$th PU band is below the threshold.
In Fig. \ref{fig:comp_EE_SE} and as expected, the SU rate is higher if perfect spectrum sensing is assumed because the transmit power is higher. Additionally, the EE (in J/bits) is higher when compared to its counterpart that considers spectrum sensing errors due to increasing the transmit power as discussed in Fig.  \ref{fig:comp_interference}.

%% file: conc.tex
\section{Conclusions} \label{sec:conc}
In this paper, we proposed an optimal power loading algorithm that optimizes the EE of an OFDM-based CR system under different channel uncertainties. The algorithm considers the channel estimation errors for the links between the SU transmitter and receiver pairs and also the effect of the imperfect sensing capabilities of the SU. Further, the algorithm does not require perfect CSI for the links from the PUs receivers to the SU transmitter. Simulation results showed that increasing the channel estimation errors deteriorates the EE. Further, they showed that assuming that the SU has perfect sensing capabilities deteriorates the EE and violates the interference constraints at the PUs receivers.
Additionally, the results demonstrated that the proposed algorithm guarantees a minimum QoS for the SU at the expense of deteriorating the EE.